# Droplet Evaporation on Porous Nanochannels for High Heat Flux Dissipation

*Sajag Poudel, An Zou, Shalabh C. Maroo\**

Department of Mechanical & Aerospace Engineering, Syracuse University, NY 13244; *corresponding author email: scmaroo@syr.edu

## Abstract

Droplet wicking and evaporation in porous nanochannels is experimentally studied on a heated surface at temperatures ranging from 35°C to 90°C. The fabricated geometry consists of cross-connected nanochannels of height 728 nm with micropores of diameter 2 μm present at every channel intersection; the pores allow water from a droplet placed on the top surface to wick into the channels. Droplet volume is also varied and a total of 16 experimental cases are conducted. Wicking characteristics such as wicked distance, capillary pressure, viscous resistance and propagation coefficient are obtained at the high surface temperatures. Evaporation flux from the nanochannels/micropores is estimated from the droplet experiments, but is also independently confirmed via a new set of experiments where water is continuously fed to the sample through a microtube such that it matches the evaporation rate. Heat flux as high as ~294 W/cm$^2$ is achieved from channels and pores. The experimental findings are applied to evaluate the use of porous nanochannels geometry in spray cooling application, and is found to be capable of dissipating high heat fluxes upto ~77 W/cm$^2$ at temperatures below nucleation, thus highlighting the thermal management potential of the fabricated geometry.

## 1. Introduction

Thin film evaporation manifests itself in nearly all evaporation processes[1-6], and surfaces are designed to amplify its occurrence to achieve high heat flux removal[2, 3, 5]. For example, over the past few years, micro/nano structures have been fabricated on surfaces to passively wick the liquid and augment thin film meniscus area, thus enhancing heat transfer.[2, 3, 5, 7-9] The way in which liquid is supplied to such structured surfaces give rise to two distinct scenarios, first where the surface is partially submerged in a pool of bulk liquid thus providing unlimited supply of liquid to the structures[10, 11], and second where liquid supply to the structured surface is limited[12] but recurs at regular intervals. An example of the latter is spray cooling[1, 13 14] where micro/macro sized droplets are dispersed, at a desired frequency, on a heated surface where the droplets wick into the structures creating thin film regions. Unlike the first scenario, such droplet coupled thin-film evaporation present unique challenges in heat transfer characterization, and is the focus of this work. Challenges arise due to the dynamic and transient nature of interaction between the thin-film menisci present within the structures with the continuously changing droplet's interfacial curvature as well as decreasing droplet volume. In particular, estimation of heat flux at the surface as well as in micro/nano structures, along with dryout limits, are important to optimize structure/liquid supply design and maximize heat flux removal. Such in-depth and complete fundamental knowledge is lacking in literature[9, 15, 16], and can advance not only thermal management solutions like spray cooling and drop-wise cooling[17] but also manufacturing related technologies such as thin-film coating[18], nano-fabrication[19], and ink-jet printing[20].

Here, we report an experimental investigation of wicking and evaporation of deionized (DI) water droplet in porous nanochannels at varying surface temperatures upto 90°C. The well-defined geometry of micropores and nanochannels help estimate the sample's porosity precisely (ε = 0.70) enabling us to determine the heat flux at these length scales. The outcome of evaporation rate from experiments is coupled with ideal spray conditions to predict surface heat fluxes which can potentially be removed via spray cooling. We also illustrate a way to maximize heat flux dissipation in spray cooling by optimum utilization of space in the porous nanochannels.

Poudel, Zou, Maroo; Syracuse University; scmaroo@syr.edu

## 2. Methods

The porous nanochannel sample comprises of cross-connected buried nanochannels of height $H$ ~ 728 nm with a micropore of diameter $d_p$ ~ 2 µm present at each interconnection (Figs. 1A and 1B). The channel width $W$ ~ 4.5 µm and spacing $S$ ~ 5.7 µm, along with $H$ and $d_p$, are indicated in the isometric view of 2×2 unit cells of the sample shown in Fig. 1B. Details of the steps of nanofabrication as well as characterization by atomic force microscopy are available in supplemental information as well as prior work by the authors[9, 21]. The micropores allow liquid from a droplet to wick into the nanochannels while the interconnected channels enable liquid exchange (a sketch with a droplet on top and simultaneous wicking is shown in Fig. 1A). Figure 1C illustrates a vertical cross-section of the heated sample with a wicked in droplet, menisci in channels and pores from where thin-film evaporation occurs, and evaporation from droplet interface. Figure 1D shows the top view of a droplet on the surface wicking into the fabricated sample of size 1.4 cm × 1.4 cm. A zoomed-in view of pores and channels is captured in Fig. 1E-1 while a goniometer image of droplet contact angle is shown in Fig. 1E-2. Further, Figs. 1F-1 and 1F-2 show a 3-D AFM image of the fabricated sample and the corresponding wall profile, respectively.

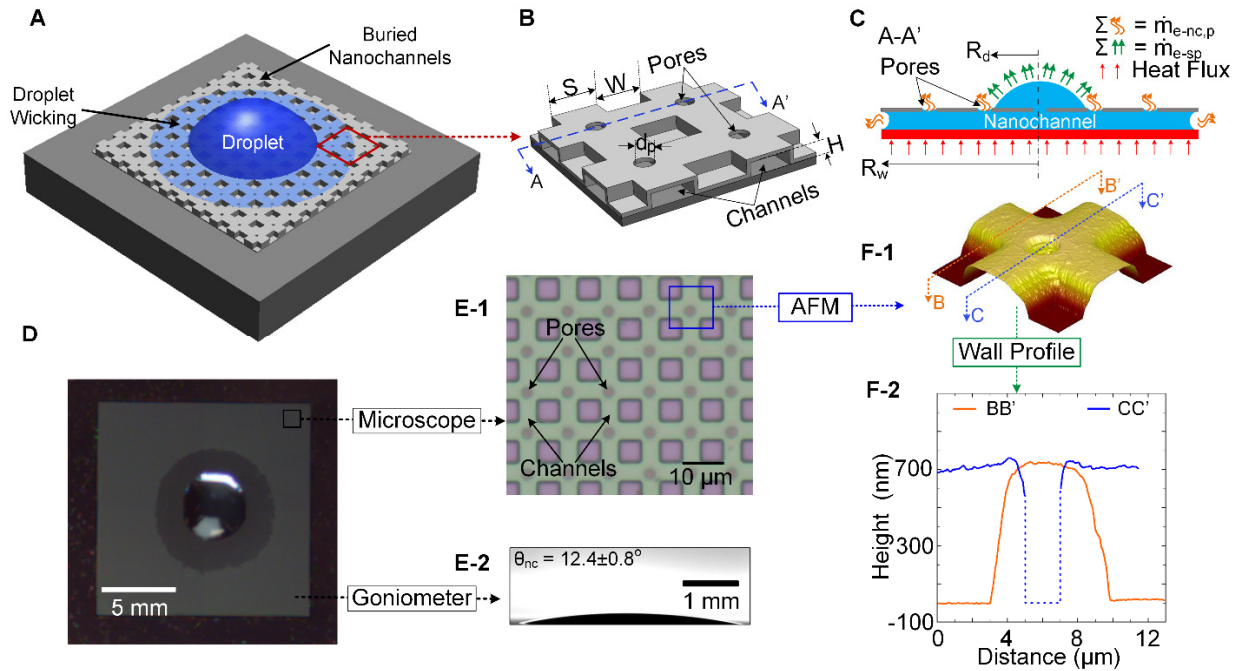

*Figure 1. Porous nanochannels sample with channel height of 728 nm and 2 µm pores. (A) Sketch of a droplet spreading and wicking into the porous nanochannels sample. (B) Isometric view of a 2×2 unit cells showing geometrical details. (C) A cross-section sketch depicting the wicking parameters and evaporation sites. (D) A high-speed camera experimental image from the top with a water droplet sitting on the sample and water wicking in the nanochannels. (E1) Top-view of the fabricated sample showing the channels and pores as observed under an optical microscope. (E2) Contact angle of the water droplet on the sample surface. (F) AFM image of a unit cell of the sample along with height profile.*

In the experimental setup, the fabricated sample is bonded atop a copper rod which has embedded cartridge heaters and is enclosed in a Teflon shell; thus, the copper rod heats the bottom surface of the sample. The temperature $T$ of the bottom surface of sample is maintained by using a PID controller while the sessile droplet of DI water is generated using a syringe pump. A high-speed camera is mounted overhead the arrangement to visualize wicking from the top while a side camera at an inclination of 12° is used to monitor the temporal variation of the amount of liquid sitting on the top. The two cameras are synchronized based on the instant when the droplet touched the surface, and the maximum error in synchronization is estimated to be 0.053 s based on the frame rates. Additional details of the experimental setup, repeatability of the experiments, and precision of measurement techniques are provided in the supplemental information. Experiments are carried out with droplets of varying volume from 4 µl to 10 µl at surface temperatures varying from 35°C to 90°C as listed in Table 1 (for e.g. C50-4 corresponds to the Case with surface temperature 50°C and droplet volume 4 µl). The droplet is placed on the porous nanochannels after the surface temperature achieved a steady state. The maximum surface temperature is limited to 90°C in order to avoid boiling in the droplet bulk liquid.

*Table 1. Different cases of wicking experiments conducted based on surface temperature and droplet volume.*

| Case | Temp. $T$ | Droplet Vol. $V$ | Case | Temp. $T$ | Droplet Vol. $V$ |
|---|---|---|---|---|---|
| C35-4 | 35 °C | 4 µl | C75-4 | 75 °C | 4 µl |
| C35-6 | | 6 µl | C75-6 | | 6 µl |
| C35-8 | | 8 µl | C75-8 | | 8 µl |
| C35-10 | | 10 µl | C75-10 | | 10 µl |
| C50-4 | 50 °C | 4 µl | C90-4 | 90 °C | 4 µl |
| C50-6 | | 6 µl | C90-6 | | 6 µl |
| C50-8 | | 8 µl | C90-8 | | 8 µl |
| C50-10 | | 10 µl | C90-10 | | 10 µl |

## 3. Results and Discussion

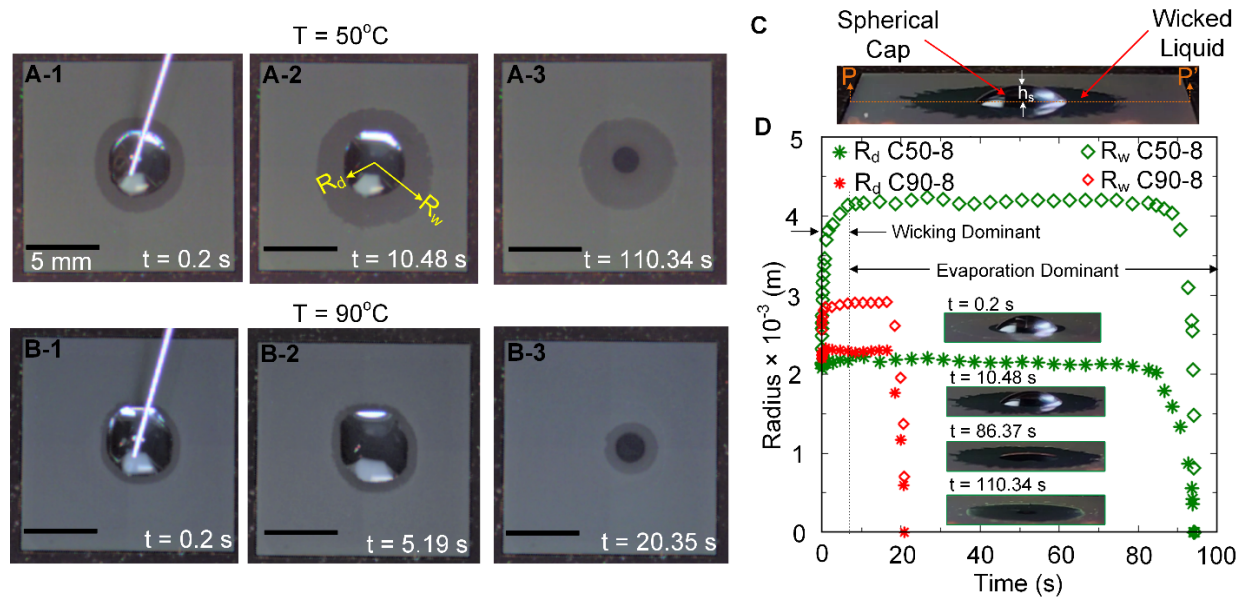

*Figure 2. Water droplet spreading and simultaneous wicking into nanochannels. High speed camera images showing the top view of the sample at surface temperature T of (A) 50°C and (B) 90°C. (C) Side camera view of droplet wicking for case C50-8. (D) Time evolution of wicking radius $R_w$ and droplet base radius $R_d$ for the two temperature cases along with inset images for case C50-8 from the side camera. Wicking dominant and evaporation dominant regimes are shown for case C50-8.*

We first study the characteristics of the liquid droplet wicking in our porous nanochannels sample at the aforementioned surface temperatures, and also explore the role of capillary pressure $P_{cap}$ and viscous resistance $K_{vr}$ on wicking dynamics at those temperatures. Figures 2A, 2B show the high-speed camera images acquired during droplet spread and simultaneous wicking in the sample for cases C50-8 and C90-

Poudel, Zou, Maroo; Syracuse University; scmaroo@syr.edu

8 respectively. The effect of temperature on the wicking distance is clear from the series of images. An image from the side camera is also shown for the instant $t$ = 10.48 s for case C50-8 in Fig. 2C. From these images, the time evolution of the wicking radius $R_w$, droplet base radius $R_d$ and wicking distance $w_d = R_w - R_d$ as well as the height of the spherical cap sitting on the top $h_s$ is acquired. Figure 2D plots the evolution of $R_w$ and $R_d$ with time for the two temperature cases (additionally, insets show images from the side camera for case C50-8). It is observed that both $R_w$ and $R_d$ increase quickly to their corresponding maximum values during the initial stage; this is the wicking dominant regime (Fig. 2D) where liquid propagation inside the nanochannels is governed by capillary and viscous forces[21]. Afterwards, the wicking front remains nearly steady as apparent from the plateau of data points in Fig. 2D; during this equilibrium phase, the evaporation rate at the menisci balances the liquid wicking rate into the channels, and is termed evaporation dominant regime. All through this regime, the droplet's spherical cap base radius $R_d$ remains steady while the droplet height $h_s$ as well as droplet's spherical cap volume gradually decrease (apparent from the insets in Fig. 2D) as liquid wicks into the channels. When the droplet completely wicks in, evaporation causes the wicking front to recede as seen with the decrease in $R_w$. Using the data represented in Fig. 2, we can quantify the evaporating menisci in micropores/nanochannels through the measurement of $R_w$ and $R_d$, and compute the total rate of evaporation from the temporal variation of spherical cap volume (obtained from $R_d$ and $h_s$) as explained next.

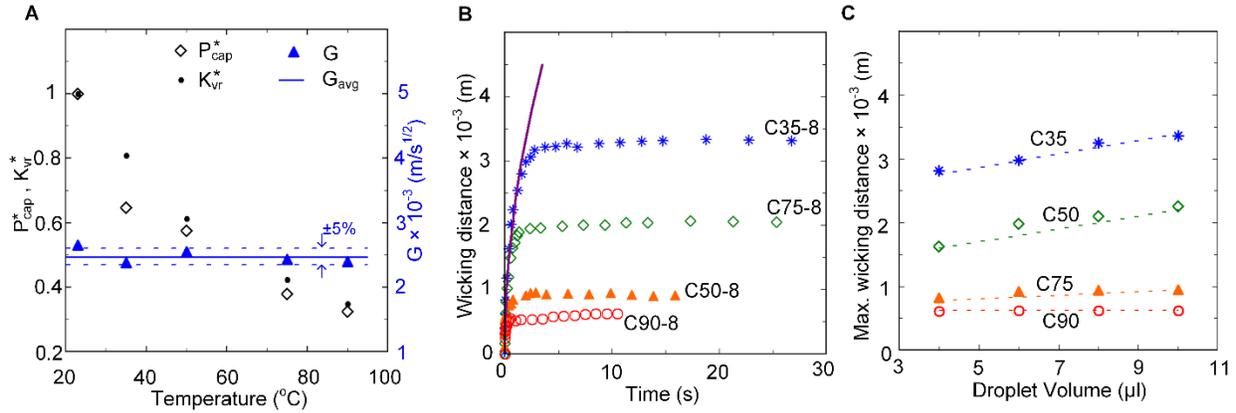

*Figure 3. Wicking characteristics of the porous nanochannels at different surface temperatures. (A) Variation of non-dimensional capillary pressure and viscous resistance, and corresponding propagation coefficient, of wicking flow inside the nanochannels in the wicking dominant regime. (B) Evolution of wicking distance in nanochannels with time. (C) Maximum wicking distance relative to droplet volume; slopes of the dotted lines (linear curve fits) are 0.11, 0.095, 0.026 and 0.0002 for C35, C50, C75, and C90 cases respectively.*

The well-defined geometry of our sample is used to determine the capillary pressure $P_{cap}$, viscous resistance $K_{vr}$ and wicking distance $w_d$ in the wicking dominant regime at different surface temperatures using the following equations: [21, 22]

$$P_{cap} = \frac{\gamma r_f cos\theta_c [2*\{(W+S)^2 - W^2\} + 4WH]}{H((S+W)^2 - S^2)} \quad \text{(Equation 1)}$$

$$K_{vr} = \frac{dP/dx}{u_{mean}} \quad \text{(Equation 2)}$$

$$w_d = G\sqrt{t} \quad \text{(Equation 3)}$$

where $\gamma$ is surface tension, $r_f$ is roughness factor of nanochannels wall ($r_f \sim 1$ from AMF images, supplementary section S1), $\theta_c$ is the intrinsic contact angle of DI water on Si substrate, $W$, $H$ and $S$ are nanochannels width, height, and spacing as explained earlier, and $G = \sqrt{2P_{cap}/K_{vr}}$ is the propagation

Poudel, Zou, Maroo; Syracuse University; scmaroo@syr.edu

coefficient.[22] Details on derivation of the equations as well as dependency of the parameters on temperature are available in supplementary information.

Figure 3A shows the variation of non-dimensional $P^*_{cap}$ and $K^*_{vr}$ with temperature in the wicking dominant regime. The non-dimensional parameters are obtained by dividing with the corresponding values at room temperature (23°C) obtained from our prior work[21] i.e. $P^*_{cap} = P_{cap}/P_{cap\_23^oC}$. As shown in Fig. 3A, both $P^*_{cap}$ and $K^*_{vr}$ decay with increase in temperature which interestingly results in a nearly constant value of propagation coefficient $G$, thus implying that the variation of wicking distance $w_d$ with time (Equation 3) is constant regardless of temperature and droplet volume in this regime. An average value of the propagation coefficient, $G_{avg}$, for the entire temperature range is computed. The finding is also confirmed in Fig. 3B which plots data for 8 µl volume droplet against a single $w_d$ curve using constant $G_{avg}$ value. This key outcome, that the initial wicking in nanochannels is independent of temperature and droplet volume, holds significant importance for high temperature applications such as pool boiling [10, 11, 23] and spray cooling[1, 13].

Beyond the initial wicking dominant phase, the variation of wicking distance with time deviates from the one predicted by Equation 3 due to evaporation coupled with limited liquid supply from the droplet spherical cap.[21] Higher surface temperatures cause early divergence of $w_d$ from the prediction of Equation 3 (Fig. 3B). Since, $G_{avg}$ only predicts the variation of wicking distance with time in the initial regime, the maximum extent ($w_{d-max}$) of liquid penetration into the nanochannels would be dependent on droplet volume and evaporation rate (thus the surface temperature). Thus, $w_{d-max}$ occurs when the wicking rate of liquid balances the evaporation rate at the menisci. Figure 3C plots the variation of $w_{d-max}$ against droplet volume $V$ for different surface temperatures. The dotted lines represent the linear curve-fit of the data points for each temperature. The slopes of the dotted lines suggest that, with increasing surface temperature, the impact of droplet volume on maximum wicking distance significantly decreases, with the slope being ~0 for surface temperature of 90°C implying $w_{d-max}$ does not depend on droplet volume at that temperature (Fig. 3C). Such an observation is again important for designing surfaces for high temperature applications.

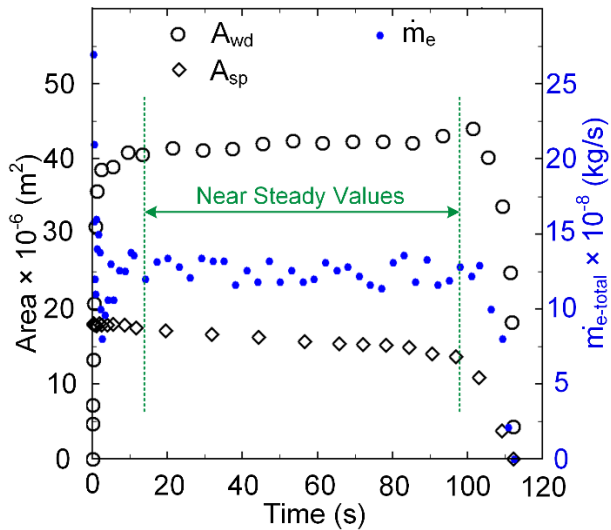

*Figure 4.* Temporal variation of wicked surface area $A_{wd}$, spherical cap area $A_{sp}$ along with the estimated total rate of evaporation $\dot{m}_{e-total}$ during droplet wicking and evaporation for case C50-8.

Next, we focus on estimating the thin-film evaporation rate occurring in the nanochannels and micropores (see Fig. 1C), along with its potential application in spray cooling. Three associated parameters are identified and determined from experimental data: wicked surface area $A_{wd}$ which only includes the area where liquid-filled micropores are exposed to ambient, droplet spherical cap surface area $A_{sp}$, and the

Poudel, Zou, Maroo; Syracuse University; scmaroo@syr.edu

total rate of evaporation $\dot{m}_{e-total}$ which is the combined evaporation rate from the menisci in nanochannels/micropores and from the droplet spherical cap interface (Fig. 1C). The temporal variation of these parameters is obtained from the data acquired from the images of the high speed camera and side camera using the following relations:

$A_{wd} = \pi(R_w^2 - R_d^2)$ (Equation 4)

$A_{sp} = \pi(R_d^2 + h_s^2)$ (Equation 5)

$V_{sp} = \frac{1}{6}\pi h_s(3R_d^2 + h_s^2)$ (Equation 6)

$\dot{m}_{e-total} = \frac{\Delta V_{sp}\rho}{\Delta t}$ (Equation 7)

The temporal evolution of the three parameters $A_{wd}$, $A_{sp}$ and $\dot{m}_{e-total}$ is shown for case C50-8 in Fig. 4. Area $A_{wd}$ increases quickly to attain a steady value with a stable wicking front within a few seconds. Area $A_{sp}$ decreases gradually due to wicking and evaporation, and can be assumed near steady in the limit 14 s < t < 98 s. Steady contact line above the nanochannels (static $R_d$) but slowly diminishing height of the spherical cap $h_s$ as observed in Fig. 2 is the reason behind this near steady value of $A_{sp}$ in this range. Interestingly, $\dot{m}_{e-total}$ also turns out to be steady within the same time range. Thus, the three parameters have near steady values in majority of the evaporation dominant regime; similar observation is made for each experimental case. Correspondingly, average values of $A_{wd}$, $A_{sp}$ and $\dot{m}_{e-total}$ are determined within their near steady range for each case, and plotted in Figs. 5A, 5B, and 5C, respectively. These three parameters are interconnected because the total rate of evaporation $\dot{m}_{e-total}$ comprises of the evaporation from the nanochannels/pores (directly related to $A_{wd}$) and the evaporation from the bulk liquid surface (spherical cap area $A_{sp}$).

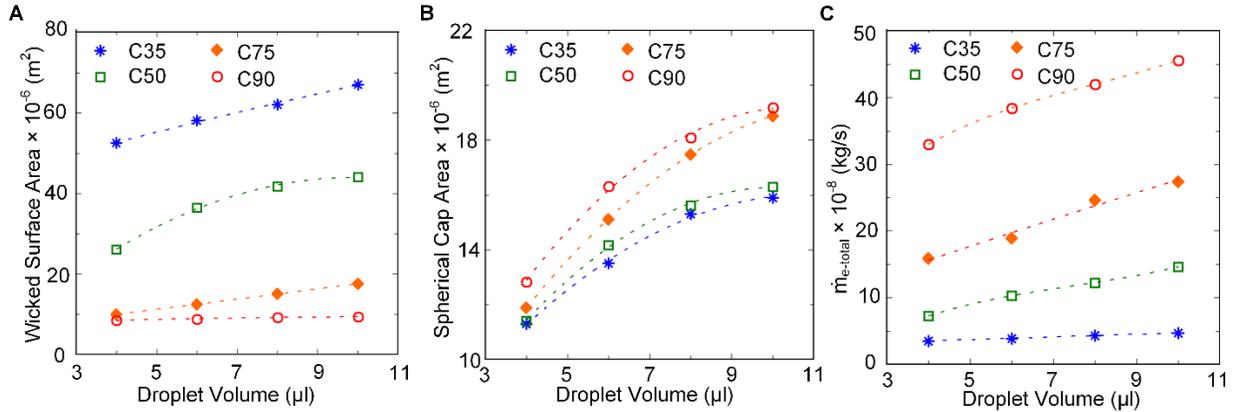

*Figure 5.* Variation of (A) wicked surface area $A_{wd}$, (B) spherical cap area $A_{sp}$, and (C) total evaporation rate variation relative to droplet volume during evaporation dominant regime for different surface temperatures. Dotted lines serve as guide for eyes.

In order to only determine evaporation flux ($\dot{m}_{e-nc,p}^{"}$) from channels/pores where thin-film menisci are present, we first decouple it from the droplet spherical cap evaporation flux ($\dot{m}_{e-sp}^{"}$) by expressing $\dot{m}_{e-total}$ as:

$\dot{m}_{e-total} = A_{sp} * \dot{m}_{e-sp}^{"} + (A_{nc} + A_p) * \dot{m}_{e-nc,p}^{"}$ (Equation 8)

where, $A_{nc}$ (= $2\pi R_w H * k_{nc}$) is the projected area of the curvatures inside nanochannels (i.e wicking front), and $A_p$ (= $A_{wd} * k_p$) is the projected area of the curvatures at micropores. The constants $k_{nc}$ and $k_p$ are based on geometry of cross connected nanochannels and micropores with values ~0.5 and ~0.0314, respectively. For a surface temperature case, Equation 8 has two unknowns, $\dot{m}_{e-sp}^{"}$ and $\dot{m}_{e-nc,p}^{"}$, which can be solved with four sets of experimental data corresponding to the four droplet volume experiments performed at that temperature. Values of $\dot{m}_{e-sp}^{"}$ and $\dot{m}_{e-nc,p}^{"}$, for each temperature, are calculated within a small range of error (<5%). Although, Equation 8 is based on the assumption of linearity with $A_{wd}$ and $A_{sp}$, we confirm this assumption as well the flux values through an additional and independent set of experiments as explained next.

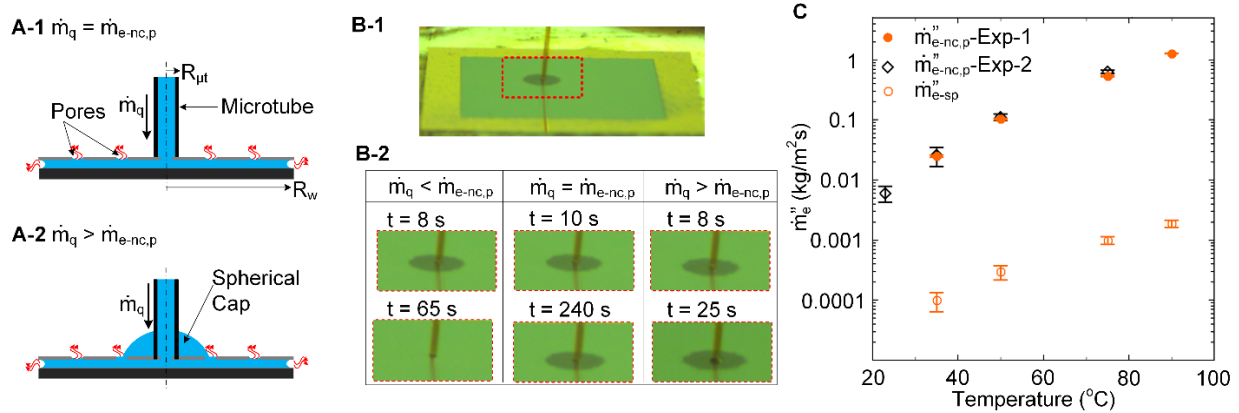

*Figure 6.* Direct experimental measurement of thin-film evaporation flux from nanochannels and micropores using continuous liquid supply. (A) Sketch to demonstrate the comparison between the flow rate of the pump $\dot{m}_q$ and the evaporation rate from nanochannels which result in the state of (A-1) equilibrium ($\dot{m}_q = \dot{m}_{e-nc,p}$) or (A-2) liquid accumulation above nanochannels ($\dot{m}_q > \dot{m}_{e-nc,p}$). (B-1) Side view of wicking and evaporation in porous nanochannels sample due to liquid supply through microtube. (B-2) Three different instances in incremental steps of $\dot{m}_q$ during the experiment on nanochannels sample at 35°C demonstrating dry out for $\dot{m}_q < \dot{m}_{e-nc,p}$, stable wicking front for $\dot{m}_q = \dot{m}_{e-nc,p}$ and liquid accumulation on top for $\dot{m}_q > \dot{m}_{e-nc,p}$. (C) Variation of thin-film evaporation flux with surface temperature from nanochannels/micropores and droplet spherical cap.

We performed a new set of experiments to directly measure the evaporation rate $\dot{m}_{e-nc,p}$ of thin-film menisci in nanochannels/micropores at same surface temperatures as before. We aim to acquire a steady-state wicking front in the heated sample such that the evaporation rate $\dot{m}_{e-nc,p}$ is balanced by the water feed rate $\dot{m}_q$ to the sample through a microtube, i.e. $\dot{m}_q = \dot{m}_{e-nc,p}$ (Fig. 6A-1). If $\dot{m}_q > \dot{m}_{e-nc,p}$, liquid would accumulate on top of the channels (Fig. 6A-2) causing additional evaporation from the accumulated bulk liquid which would introduce error in $\dot{m}_{e-nc,p}$ measurement. Thus, in the experiments, flow of DI water through the microtube and onto the sample is controlled by an automatic-pump, and tailored such that accumulation of bulk liquid atop the surface is avoided (Fig. 6B-1). Such a delicate flow balance is achieved by gradually increasing the liquid flow rate $\dot{m}_q$ of the pump in small steps starting from $\dot{m}_q = 0$ kg/s (= 0 μl/min). Figure 6B-2 shows three different instances of gradual increment of $\dot{m}_q$, for the surface temperature of 35°C. At small flow rate of $\dot{m}_q$ (first column of Fig. 6B-2), wicking is initially observed but the wicked liquid dries out over time as the liquid flow rate is lower than the rate of evaporation i.e. $\dot{m}_q < \dot{m}_{e-nc,p}$. With increase in flow rate $\dot{m}_q$, we obtain a stable wicking front when the evaporated liquid from the nanochannels sample is exactly balanced by the liquid supplied through the microtube ($\dot{m}_q = \dot{m}_{e-nc,p}$) as seen in the second column. Further increment in $\dot{m}_q$ ($> \dot{m}_{e-nc,p}$) results in an undesirable accumulation of liquid on top of the channels as seen in the third column. After repeating similar set of experiments for all temperatures, we achieve the steady state thin-film evaporation rate $\dot{m}_{e-nc,p}$ from the menisci in the nanochannels and micropores.

Poudel, Zou, Maroo; Syracuse University; scmaroo@syr.edu

Considering the projected area of menisci inside nanochannels and micropores during evaporation, we obtain the evaporation rate flux $\dot{m}''_{e-nc,p}$ as follows:

$$\hat{A}_{wd} = \pi\left(\hat{R}_w^2 - \hat{R}_{\mu t}^2\right) \quad \text{(Equation 9)}$$

$$\dot{m}''_{e-nc,p} = \frac{\dot{m}_{e-nc,p}}{\hat{A}_p + \hat{A}_{nc}} = \frac{\dot{m}_{e-nc,p}}{\hat{A}_{wd}*k_p + 2\pi\hat{R}_w H*k_{nc}} \quad \text{(Equation 10)}$$

where, $\hat{R}_w$ and $\hat{R}_{\mu t}$ are the radii of wicked liquid and microtube respectively (Figure 6A-1). The variation of $\dot{m}''_{e-nc,p}$ with temperature obtained from these set of experiments is plotted in Figure 6C (labelled as $\dot{m}''_{e-nc,p}$-Exp-2). Figure 6C also shows the evaporation flux (labelled as $\dot{m}''_{e-nc,p}$-Exp-1) obtained through droplet wicking experiments as discussed earlier i.e. the values of $\dot{m}''_{e-nc,p}$ and $\dot{m}''_{e-sp}$ obtained by solving Equation 8. As can be seen, the values of $\dot{m}''_{e-nc,p}$ obtained from the two independent experiments are in excellent agreement. The droplet experiments also allow us to estimate evaporation flux from the droplet interface $\dot{m}''_{e-sp}$ (Fig. 6C) and we find it to be about two orders of magnitude smaller than $\dot{m}''_{e-nc,p}$. Thus, thin-film evaporation and its augmentation is highly desired in droplet based thermal management solutions such as spray cooling.

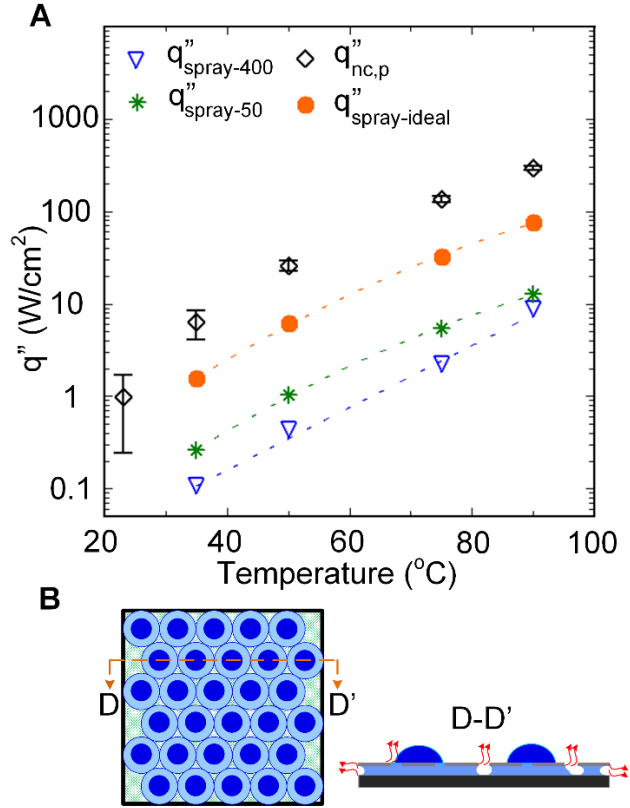

*Figure 7. Potential use of porous nanochannels in spray cooling thermal management for high heat flux dissipation (A) Variation of heat flux with surface temperature based on the projected area of thin-film menisci present in channels/pores from evaporation experiments, along with the cooling performance (based on projected area of the sample) using different spray parameters. (B) A sketch demonstrating fcc distribution of droplets and corresponding wicking in nanochannels to achieve high heat flux dissipation.*

Next, we demonstrate the potential of using our porous nanochannel design in spray cooling to achieve high heat flux dissipation. Using the experimentally measured evaporation flux rate $\dot{m}''_{e-nc,p}$, we calculate the heat flux in the nanochannels and micropores using the relation:

$$\dot{q}''_{nc,p} = \dot{m}''_{e-nc,p} * h_{fg} \qquad \text{(Equation 11)}$$

where $h_{fg}$ is the latent heat of evaporation at the corresponding surface temperature. The obtained variation of heat flux in channels/pores is plotted against temperature in Figure 7A. Heat flux as high as ~294 W/cm$^2$ at surface temperature of 90°C is achieved through thin-film evaporation. In order to extend the present fundamental work to a relevant practical application of thermal management, we examine the performance of idealized spray cooling on the fabricated porous nanochannels sample and estimate maximum possible heat flux removal rates. As shown in Figure 7B, we consider an ideal face-centred-cubic (fcc) distribution of the wicked surface area of sprayed droplets atop the heated sample. Two cases are investigated where different spray-generated droplet diameters of 400 μm and 20 μm (represented as spray-400 and spray-20, respectively), based on available literature and practical feasibility[14, 24], are assumed to be deposited on the sample. From the volume of the individual droplet of spray $V_{d-s}$ and the contact angle of the liquid on the nanochannels sample $\theta_{nc}$, its corresponding spherical cap base radius $R_{d-s}$ is evaluated and the corresponding wicking radius $R_{w-s}$ is estimated from the existing results of the wicking test (see supplemental information). Such an analysis helps estimate the maximum number of spray droplets sitting in fcc arrays on our sample (see Figure 7B). Using the droplet distribution, the projected area of menisci in micropores and nanochannels is obtained from which the spray cooling performance $\dot{q}''_s$ (heat flux removal based on the projected area of the sample) for each set of spray at the corresponding temperature is calculated:

$$\dot{q}''_s = \frac{\dot{q}'' * (A_{p-s} + A_{nc-s})}{A_{sample}} = \frac{\dot{q}'' * n_d \left( \pi (R_{w-s}^2 - R_{d-s}^2) * k_p + 2\pi R_{w-s} H * k_{nc} * k_{nc} \right)}{A_{sample}} \qquad \text{(Equation 12)}$$

where, $n_d$ is the number of droplets in the array obtained sitting atop the channels based on fcc distribution and a sample size of 1.4 cm × 1.4 cm ($A_{sample}$ = 1.96 cm$^2$). For the two sprays, spray-400 and spray-20, the values for $n_d$ are found to be in a wide range of 15-96 and 6,000-380,000 respectively (see Table S1 in supplemental information) at different temperatures. Figure 7A shows the estimated heat flux removal for the two sprays $\dot{q}''_{spray-400}$ and $\dot{q}''_{spray-20}$; high heat flux dissipation of ~12.80 W/cm$^2$ can ideally be attained.

Additionally, in Figure 7A, we observe that the extent of heat flux removal increases significantly (Y-axis being logarithmic), with a smaller droplet size of spray. Also, we know from the wicking test at high temperature, the proportion of the spherical cap volume from the total droplet volume decreases for smaller droplet volume (see supplemental information). Thus, we can estimate a critical limit of droplet volume for wicking in nanochannels such that, the proportion of the spherical cap is zero (i.e. no spherical cap) and the entire liquid is wicked in. This critical limit is achieved when the droplet has a diameter ~4.5 μm. Hence, we again perform a similar spray cooling analysis but with an ideal spray of uniformly generated droplet diameters of ~4.5 μm. The variation of heat flux removal achieved through this ideal spray is also plotted in Fig. 7A. The heat flux dissipation can be significantly increased to ~77 W/cm$^2$ at substrate temperature of 90°C with such small droplet diameters. It is noteworthy to mention here that for the identical nanochannels sample, we had reported pool boiling heat flux based on the projected area to be ~20 W/cm$^2$ at a surface temperature of 117°C, and critical heat flux ~178 W/cm$^2$ at a surface temperature of ~140°C.[11] Thus, proper liquid supply design (e.g. injecting liquid or use of multiple spray nozzles) in our porous nanochannel surfaces can enable high heat flux dissipation similar to that achieved in pool boiling heat transfer while avoiding the complexity of a boiling setup. This foregoing analysis highlights the thermal management potential of using thin-film evaporation in such structured surfaces via spray cooling.

## 4. Conclusion

A well-defined porous geometry of cross-connected nanochannels with micropores was fabricated and heated to study droplet coupled thin-film evaporation. A droplet placed on the surface wicks into the channels through the pores and enables tracking of the evaporating menisci through time-resolved visualization. For various surface temperatures ranging from 35°C – 90°C and varying droplet volumes from 4 µl to 10 µl, wicking characteristics and evaporation rate were determined. It was found that initial wicking (i.e. in wicking dominant regime) in the nanochannels was independent of surface temperature and droplet volume. In the later stage of droplet wicking, i.e. evaporation dominant regime, the maximum wicking distance does not depend on droplet volume at high surface temperatures. Evaporation flux from channels/pores, where the thin-film menisci are present, was found to be about two orders of magnitude higher than from the droplet interface, and corresponding heat flux as high as ~294 W/cm$^2$ was obtained from the channels/pores. Applying the experimental findings of heat transfer and wicking characteristics towards potential use of spray cooling based thermal management, high heat flux dissipation ~ 77 W/cm$^2$ can ideally be attained from thin-film evaporation in the porous nanochannels. Additional enhancements in cooling performance can be achieved by further optimizing the nanochannels/micropores geometry.

## 5. Supplemental Information

Details on nanofabrication of the sample, experimental setup and procedure, analytical models, and spray droplet parameters are available in supplemental information.

## 6. Acknowledgement

The material presented in this research is based on the work supported by the Office of Naval Research under contract/grant no. N000141812357. This work was performed in part at Cornell NanoScale Facility, an NNCI member supported by NSF grant NNCI-2025233.